\documentstyle[12pt]{article}

\begin{document}
\thispagestyle{empty}
\setcounter{page}0

\newcommand {\be}{\begin{equation}}
\newcommand {\ee}{\end{equation}}
\newcommand {\bea}{\begin{array}}
\newcommand {\cl}{\centerline}
\newcommand {\eea}{\end{array}}
\renewcommand {\theequation}{\thesection.\arabic{equation}}
\renewcommand {\thefootnote}{\fnsymbol{footnote}}
\newcommand {\newsection}{\setcounter{equation}{0}\section}

\def\nc{noncommutative }
\def\com{commutative }
\def\ncy{noncommutativity }
\def\repr{representation }
\def\reps{representations }
\def \simlt{\stackrel{<}{{}_\sim}}
\def \simgt{\stackrel{>}{{}_\sim}}
\baselineskip 0.65 cm


\begin{flushright}
IC/2000/130\\
hep-th/0009037
\vskip 14mm
\end{flushright}

\begin{center}
{\Large{\bf{Pair Production by a Constant External Field in Noncommutative
QED}}} 

{\bf\large{{N. Chair$^{\dagger}$ and M.M. Sheikh-Jabbari}}}

\vskip 12mm

{\it The Abdus Salam International Center for Theoretical Physics, 
Strada Costiera 11,Trieste, Italy}\\
E-mail: {\tt jabbari@ictp.trieste.it} \\

$^{\dagger}$ {\it Physics Department, Al-alBayt University, 
Mafraq-Jordan}\\
E-mail: {\tt N.Chair@rocketmail.com}

\vskip 8mm

{\bf abstract} 

\end{center}
In this paper we study QED on the \nc space in the constant electro-magnetic 
field background. Using the explicit solutions of the \nc version of Dirac
equation in such background, we show that there are well-defined 
in and out -going asymptotic states and also there is a causal Green's
function. We calculate the pair production rate in this case. 
We show that at tree level noncommutativity will not change
the pair production and the threshold electric field. 
We also calculate the pair production rate considering the first loop
corrections. In this case we show that the threshold electric field is
decreased by the noncommutativity effects.

\newpage

\section{Introduction}
\setcounter{equation}{0}

Noncommutative spaces naturally come about when one studies D-brane worldvolume 
theory in the B-field background \cite{SW}. Such \nc spaces can be characterized
by the coordinate operators, ${\hat X}^{\mu}$, satisfying
\be\label{NC}
[{\hat X}_{\mu},{\hat X}_{\nu}]=i\theta_{\mu\nu},
\ee
where $\theta_{\mu\nu}$ is the \nc parameter and is of dimension of (length)$^2\ $.
The field theory formulated on these spaces, the \nc field theory, is then described by 
field operators which are functions of ${\hat X}$. However, using the Weyl-Moyal correspondence
\cite{{AlW},{Micu},{Ihab}} one can show that instead of operator valued functions, it is enough to take the
corresponding \com theory and replace the product of fields by the {\it star} product:

\be\label{star}
(f*g)(x)=exp({i\over 2}\theta_{\mu\nu}\partial_{x_{\mu}}\partial_{y_{\nu}}) 
f(x)g(y)|_{x=y},
\ee
where $f$ and $g$ are two arbitrary functions and assumed to be
infinitely differentiable. Then, in order to quantize the theory one should specify the 
Hilbert space (or equivalently the measure, in the path integral
formulation). Thanks to the properties of star product under the integral
over the space-time, star product
in the quadratic terms of the action can be removed, suggesting that the
asymptotic states of free
field theory can be consistently chosen the same as the corresponding
\com theory; i.e. perturbative
Hilbert space of a \nc field theory is that of the \com one \cite{Micu}.
However, this argument is true
if, the conjugate momentum of the field is the same as its \com
counter-part, for which the definition of 
star product (\ref{star})  implies that the time components of
$\theta_{\mu\nu}$, $\theta_{0i}$,
should be zero and hence, we will restrict ourselves only to these
cases. In fact it has been shown that
theories with time-like \ncy ($\theta_{0i}\neq 0\ ,\ \theta_{ij}=0$) are not unitary \cite{Mehen}, however 
those with light-like \ncy lead to well-defined quantum theories \cite{Seib}.

Here, we investigate some more possible phenomenological aspects of \nc field theories, and in
particular \nc
version of QED. Some other phenomenological consequences of \nc standard model have
been addressed in \cite{Roiban}. So, first we start with introduction of
pure \nc gauge theories, then we add fermions and as
usual, form of the interaction terms are fixed using the \nc gauge
invariance. 

Performing explicit calculations of two and three point functions, it has been shown that NCQED satisfies
on-shell Ward identity and is renormalizable (at one loop) \cite{{Ihab},{Haya},{Martin}}. Studying effective
interaction vertex it has been shown that \cite{Ihab}:

{\it i)} magnetic dipole moment of a \nc Dirac particle at one loop level has a {\it 
spin independent} part which is directly proportional to \ncy parameter, $\theta$; 

{\it ii)} any moving  \nc Dirac particle shows {\it electric dipole} effects, which of course receives quantum
corrections at one loop.

The behaviour of NCQED under discrete symmetries (C, P and T) have also been addressed and shown that for
space-like \ncy ($\theta_{0i}=0$) the theory is parity preserving while CP violating. More precisely, under
charge conjugation NCQED on $R^4_{\theta}$ is mapped into a NCQED on $R^4_{-\theta}$. Also it has been
shown that the theory preserves CPT \cite{CPT}.
Hence it is plausible to look for some generalization of the usual CPT theorem which is also true
for the Lorentz non-invariant cases like the \nc Moyal plane \cite{prog}.

In this paper we study pair production in NCQED by a constant
electro-magnetic field background. The constant background has been 
previously considered in \cite{AlB}. 
The peculiar feature of the constant electro-magnetic field is
the appearance of gauge non-invariant quantities. This is due to the fact
that the "trace" which makes the operators to be gauge invariant, in the
\nc case is replaced by the integration over the space-time, and besides the
integration one should also neglect the surface terms at infinity, which
of course is not true for the constant field strength case \cite{AlB}.

To find the pair production rate, we use Nikishov
method which is based on explicit solutions of the corresponding Dirac
equation \cite{{Niki1},{Niki2}} (for a more
recent review and more detailed references see also \cite{Krug}).

Solving the \nc Dirac equation in the constant electro-magnetic field
background we show that, similar to the \com case, there are well-defined
in and out going fermionic states and hence, there is also a well-behaved
Green's function. Having the proper states and propagator we work out the
rate of pair production in the unit volume, unit time.  
In the tree level, we show that the \nc effects do not appear in the physical rate.
As a result, at tree level, the threshold electric field for the pair production is not affected
by \nc corrections. 

In addition we perform the calculations considering the first loop effects
on the magnetic dipole moment of electron in NCQED \cite{Ihab}. In this
case we show that the threshold electric field receives some corrections
due to noncommutativity.  

\vskip .5cm
\section{NCQED, the action}
\setcounter{equation}{0}
In order to get the \nc Dirac equation coupled to NCU(1) theory, we first
build the action:

{\it i)Pure Gauge theory}
\newline
The action for the pure gauge theory is
\be\label{AA}
S={1\over 4\pi}\int F_{\mu\nu}*F^{\mu\nu}d^4x=
{1\over 4\pi}\int F_{\mu\nu}F^{\mu\nu}d^4x\ ,
\ee
with 
\be
F_{\mu\nu}=\partial_{[\mu}A_{\nu]}+ie \{A_{\mu},A_{\nu}\}_{MB}\ .
\ee
In the above $e$ is the gauge coupling constant and the $MB$ stands for Moyal bracket defined as
$\{f,g\}_{MB}=f*g-g*f\ $. One can show
that the above action enjoys the 
\nc gauge transformations\cite{{SW},{Haya}}

$$
A_{\mu}\rightarrow A'_{\mu}= 
U(x)*A_{\mu}*U^{-1}(x)+ {i\over g}U(x)*\partial_{\mu}U^{-1}(x)\ ,
$$
\be\label{UA}
U(x)=exp*(i\lambda),\;\;\;\;\ U^{-1}(x)=exp*(-i\lambda)\ ,
\ee
where
\be\bea{cc}
exp*(i\lambda(x))\equiv 1+i\lambda-{1\over 2} \lambda*\lambda-{i\over 3!}
\lambda*\lambda*\lambda+...\ , \\
U(x)*U^{-1}(x)=1\ .
\eea\ee
Since here we are only interested in NCQED, we choose $\lambda,\ A_{\mu}$ to be in
$U(1)$ algebra. However, this can easily be extended to $U(n)$ valued functions giving rise to NCU(n) theory. 

{\it ii)Fermionic Part}
\newline
Fermions can be added to the above gauge theory, developing the definition 
of "covariant derivative". In the
NCQED, it has been shown that there are two different kinds of covariant derivatives related by charge
conjugation. In other words, there are two different types of fermions which are mapped into each
other by charge conjugation, hence they can be called positively or negatively charged fermions
\cite{{Haya},{CPT}}. The explicit form of the covariant derivative for the positively charged particles
is   
\be
D^+_{\mu}\psi(x)\equiv \partial_{\mu}\psi(x)-ie(A_{\mu}*\psi)(x)\ ,
\ee
while for the particles with the negative charge it is
$$ 
D^-_{\mu}\psi(x)\equiv \partial^{\mu}\psi(x)+ie(\psi*A_{\mu})(x)\ .
$$
In this paper we only consider the $D^+$ case, and the other, $D^-$ can be recovered by just sending
$\theta$ to $-\theta$. The fermionic part of NCQED action is then
\be
S_f=\int d^4x \bar\psi*(-i\gamma^{\mu}D^+_{\mu} -m)\psi\ .
\ee
It is easy to verify that this action is also invariant under NCU(1) transformation defined by 
$\psi\to U*\psi$ and (\ref{UA}).
Using the definition of the covariant derivative one can verify that:
\be
\{D_{\mu}^{\pm}, D_{\nu}^{\pm}\}_{MB}=\mp ie F_{\mu\nu}\ .
\ee 
 
\section{Pair production amplitude}
\setcounter{equation}{0}

To find the pair production rate, we solve the \nc Dirac equation, 
\be
(-i\gamma^{\mu}\partial_{\mu}-m)\psi+\ e\gamma^{\mu} A_{\mu}*\psi=0\ ,
\ee
in the constant field background. Since there is a Lorentz transformation which maps constant 
electric and magnetic field ($\vec{E}$ and $\vec{B}$) into parallel $\vec{E}$ and
$\vec{B}$ \footnote{However, we note that this is not true for the cases
in which both Lorentz-invariants, namely $E^2-B^2$
and $E\cdot B$, are zero. Hence our arguments is not covering those
cases.}, we only consider the parallel electric and magnetic fields here.
We choose our $x^3$ axis
to be along $\vec{E}$. Then the
corresponding $A_{\mu}$ field can be taken as
\be
A_{\mu}=(0,\ 0,\ Bx_1,\ -Et)\ ,
\ee
where $t$ is the time coordinate ($x^0$). Inserting this $A_{\mu}$ into the Dirac equation we obtain  
\be
(\gamma^{\mu}\Pi_{\mu}-m)\psi=0\ ;
\ee
\be\left\{\bea{cc}
\Pi_{\mu}=-i\partial_{\mu}+eA_{\mu}\ \ \ \ \ \ \ \mu\neq 2\ ,\\
\Pi_{2}=-i\partial_{2}+eBx_1+{i\over 2}eB\ \theta_{1j}\partial_j\ .
\eea\right.
\ee
Following the lines of \cite{{Niki1},{Niki2}}, it is more convenient to use the squared Dirac equation (for our
conventions see the Appendix)
\be\label{Psi}
(\Pi^2+S-m^2)\Psi=0\ ,
\ee
where
\be
S=-{ie\over 2}\gamma^{\mu}\gamma^{\nu}F_{\mu\nu}\ ,
\ee
and with our choice the non-zero components of $F$ are $F_{30}=-F_{03}=E,\ F_{12}=-F_{21}=B\ $. 
Then the fermion field, $\psi$, can be obtained as
\be
\psi=(m-\Pi_{\mu}\gamma^{\mu})\Psi\ .
\ee

In order to solve (\ref{Psi}), first we find the eigenvectors and eigenvalues 
of matrix $S$:
\be
S\Gamma_i=s_i\Gamma_i\ ,\;\;\;\;\;\; i=1,2,3,4\ .
\ee
For our purpose it is enough to consider only two of these, e.g. $i=1,2$, which hereafter we will denote them
by $+$ and $-$:
\be\label{spm}
s_{\pm}=\pm eB-ieE\ ,
\ee
and 
$$
\Gamma_{+}=(1,\ 0,\ 1,\ 0)\ , \;\;\;\;\;  \Gamma_{-}=(0,\ 1,\ 0,\ -1)\ .
$$  
Then, $\Psi$ can be decomposed into the matrix part which is proportional to $\Gamma_{\pm}$ and
the functional part
\be
\Psi_{\pm}=Z_{\pm}(x)\ \Gamma_{\pm}\ ,
\ee
where $Z_{\pm}$ satisfy the equations
\be\label{SQD}
(\Pi^2+s_{\pm}-m^2)Z_{\pm}=0\ .
\ee

Noting that $x^2,\ x^3$ do not appear in the equation (\ref{SQD}), we use the following ansatz for $Z_{\pm}$,
\be
Z_{\pm}=N\ exp(ip_2x_2+ip_3x_3)\ F_{\pm}(x_1,t)\ .
\ee
Plugging this ansatz into (\ref{SQD}), defining
\be\label{ppp}
P_2=p_2-{1\over 2}eB\ \theta_{1j}p_j\ ,
\ee
\be\label{ttt}
eE\tau^2=(p_3-eEt)^2\ ,
\ee
\be\label{xxx}
eB\rho^2=(P_2+eBx_1)^2\ ,
\ee
and separating the variables, i.e.
$$ 
F_{\pm}(\rho,\tau)=\chi_{\pm}(\rho)\ \Phi_{\pm}(\tau)\ ,
$$
we get
\be\label{Xi}
({\partial^2\over \partial\rho^2}- \rho^2 \pm 1)\chi_{\pm}=K_{\pm} \chi_{\pm}\ , 
\ee
\be\label{Phi}
({\partial^2\over \partial\tau^2}+ \tau^2+ i+ \lambda_{\pm})\Phi_{\pm}=0\ ,
\ee
where $K_{\pm}$ are constants and
\be
\lambda_{\pm}={m^2-eBK_{\pm}\over eE}\ .
\ee

Eq.~(\ref{Xi}) is basically the Schroedinger equation for a harmonic
oscillator, hence:
\be 
K_+=2(n+1)\;\;\ ,\;\;\ K_-=2n\ ,\;\;\ n=0,1,2,...\ ,
\ee
and $\chi_{\pm}$ can be written in terms of Hermite polynomials: 
$
\chi_{\pm}=e^{-{1\over 2}\rho^2}\ H_n(\rho)\ .
$
Solutions of  eq.~(\ref{Phi}) are the parabolic-cylinder functions \cite{Handbook}
\be\label{Para}
\left\{\bea{cc}
\Phi_{1+}=D_{\nu}(-(1-i)\tau) \\
\Phi_{2+}=D_{\nu *}((1+i)\tau) \eea\right. ;
\left\{\bea{cc}
\Phi_{1-}=D_{\nu *}(-(1+i)\tau) \\
\Phi_{2-}=D_{\nu}((1-i)\tau) \eea\right. \ ,
\ee
where
\be
\nu={1\over 2}(1+i\lambda_{\pm})\ .
\ee
Not all the above mentioned solutions, (\ref{Para}), are linearly independent; $\Phi_{i+}$ and
$\Phi_{i-}$
form two complete sets.
>From the asymptotic expansions of $D_{\nu}$ functions we observe that our $\Phi_{i\pm}$ are
leading to some 
well-defined states as $t\to \pm\infty$. Actually, $\Phi_{i+}$ are those
with positive frequency solutions and
$\Phi_{i-}$  with negative frequency, as $t\to \infty$. These solutions
are related through
\be
\left\{\bea{cc}
\Phi_{1+}=\beta \Phi_{2-}+ \zeta \Phi_{2+}   \\
\Phi_{2+}=\beta^*\Phi_{1-}+ \zeta^* \Phi_{1+}\ , 
\eea\right.
\ee  
with $\beta=e^{i\pi\nu}$, and $-|\beta|^2+|\zeta|^2=1$. Now that we have the final fermionic
solution, the absolute probability for pair production is then, $|\beta|^2=e^{-\pi\lambda}$. As 
it is seen the effects of
\ncy has been totally disappeared in the final result but, still one should work out the pair
production rate per unit volume, unit time. The family of our fermionic solutions
is characterized by quantum numbers $P_2,p_3,\ n$ and also $\pm$ sings, corresponding to two spin
states. Hence to find the full rate one should sum over all these set of quantum numbers. In
order to normalize our
states (and also regularize our calculations) let us put our system in a
box of sides $L$, then the average number of pairs produced is
\be
{\bar{\cal N}}=\int dP_2dp_3\sum_{n}(|\beta_+|^2+|\beta_-|^2)\ {L^2\over (2\pi)^2}\ .
\ee
Since there is no explicit $P_2,p_3$ dependence in $\beta$'s, 
using  eqs.~(\ref{ttt}), (\ref{xxx}), integration over
$p_3$ and $P_2$ can be replaced by \cite{{Niki1},{Krug}}
$$
\int dp_3 \longrightarrow eET\ \;\;\;\; ,\;\;
\int dP_2 \longrightarrow {eBL} \ ,
$$
So altogether the pair production rate, ${{\bar{\cal N}}\over L^3 T}$, is
\be\label{rate}
I_0(E,B)={\alpha EB\over\pi}\ exp(-{\pi m^2\over eE})\ \coth({\pi B\over E})\ ,
\ee
which is exactly the same as the \com results of \cite{{Niki2},{Krug}}.

Here we should remind that in all the above manipulations, instead of the eigen-values of
${\partial\over\partial x_2}$ operator, $p_2$, we have used the $P_2$, which is the eigen-value
for the "physical" momenta along $x_2$ when we have a non-zero $\theta$ parameter. This
"physical" quantum number is actually counter-part of the $x'$-coordinate system introduced in
\cite{AlB} and is not invariant under the NC$U(1)$ gauge
transformations. Then, as explained in \cite{AlB}, the extra factor arising in  going
from $p_2$ to $P_2$ frame can be absorbed in the normalization factor,
$N$, in Eq.(3.12).
Although we have not presented here the result (\ref{rate}) is invariant under the NC$U(1)$
gauge transformations defined by (2.3) and (2.5), and $E$ and $B$ are the electric and
magnetic components of (2.2).

All of the above calculations are done in the classical level. The
quantum (loop) effects can also be included if we consider the anomalous
magnetic moment of Dirac particles \cite{{gfactor},{Nouvo}}. In the
usual QED, this can easily be done by replacing $s_{\pm}$, (\ref{spm}),
with $\pm {g\over 2}eB-ieE$, where $g$ is the gyro-magnetic
ratio. However, for the NCQED case as shown in \cite{Ihab}, at one loop
level Dirac particle will also show a spin independent magnetic moment.
Taking this into account, the proper $s_{\pm}$ for the NCQED at one loop
level
is
\be
s_{\pm}=\pm {g\over 2}eB-ieE+\ {e\alpha\gamma_{{E}}\over
3\pi}m^2{\vec{\theta}.\vec{B}}\ ,
\ee
where 
\be
g-2={\alpha\over \pi} \;\;\;\ {\rm and} \;\;\; \theta_i\equiv\epsilon_{ijk}\theta_{jk}\ ,
\ee 
and $\gamma_E=\gamma_{{\rm Euller}}$.
Inserting these values for $s_{\pm}$ and repeating all the computations, 
the pair production rate up to first loop is
\begin{eqnarray}\label{Qrate}
I_{{\rm 1st\ loop}} =
{\alpha EB\over\pi}\ 
exp \big[-{\pi m^2\over eE}(1-{e\alpha \gamma_{{\rm E}}\over
3\pi}\vec{\theta}.\vec{B})\big]\times 
{\cosh({\pi g B\over 2E})\over\sinh({\pi B\over E})}\ .
\end{eqnarray} 

The interesting point is that,
the threshold electric field is reduced by the \ncy effects. 
This possible change in the threshold electric field due to \ncy can have
important astrophysical and cosmological consequences, where we have a very strong magnetic
field, e.g. for the neutron stars. This change in the pair production rate can be used to put
some (lower) bound on $\theta$. 

For the pair annihilation without photon emission amplitude, since our
theory is $T$ violating, it is not the same as pair production rate.
However, using the CPT invariance of the theory \cite{CPT}, this amplitude
is related to that of the pair production by $\theta\to -\theta$
transformation. 

The other comment we should make is that, here we only present the
calculations for spin one-half particles.  However, all of our discussions
through the lines of \cite{Krug} can be generalized to particles with
arbitrary spin. As it is expected, at classical level, we obtain the
\com results, however at the one loop  we expect to see the \nc effects. These effects are
presumably the same as (3.27) but now the factor ${1\over 3\pi}$ in the term
proportional to $\theta.{\vec B}$ is replaced by the proper numeric factor.

\vskip .3cm

{\bf Acknowledgements} \vskip.3cm

N. C. would like to thank ICTP for hospitality where this work was done.
The authors would also like to thank the anonymous referee for useful
comments. The work of M.M. Sh-J.  was partly supported by the EC contract
no. ERBFMRX-CT 96-0090.

\vskip .3cm

{\large{\bf Appendix$\;\;$ :} The $\gamma$ matrix conventions}

In this paper we used the following conventions:
$$
\gamma^0=\left( \matrix{1 & 0 \cr 0 & -1}\right)\;\;\;\; ,
\;\;\;\;\;\gamma^i=\left(\matrix {0 & {\bf \sigma}^i \cr
{\bf-\sigma}^i & 0 }\right)\ ,
$$
which lead to 
$$
\eta_{\mu\nu}={\rm diag}(+1,\ -1,\ -1,\ -1)\ .
$$

\end{document}